\documentstyle[aps,twocolumn]{revtex}
\begin{document}
\draft
\title{Single Molecule Magnetic Resonance and Quantum Computation}
\author{Haiqing Wei and Xin Xue}
\address{
Department of Electrical and Computer Engineering, McGill
University \\ Montreal, Quebec, Canada H3A 2A7 \\
{\rm dhw@tsp.ece.mcgill.ca} \\
{\rm xxue@photonics.ece.mcgill.ca}}
\maketitle
\begin{abstract}
It is proposed that nuclear (or electron) spins in a trapped molecule
would be well isolated from the environment and the state of each spin
can be measured by means of mechanical detection of magnetic resonance.
Therefore molecular traps make an entirely new approach possible for
spin-resonance quantum computation which can be conveniently scaled up.
In the context of magnetic resonance spectroscopy, a molecular trap
promises the ultimate sensitivity for single spin detection and an
unprecedented spectral resolution as well.
\end{abstract}
%03.67.Lx Quantum computation
%39.90.+d Other instrumentation and techniques for
%                     atomic and molecular physics
%89.80.+h Computer science and technology

A quantum computer (QC), if ever functions, will outperform any classical
computer in certain computation and simulation tasks; however building a
QC which is sizable and robust against decoherence imposes a great
challenge on experimental physics, even on principle design$^1$. In this
letter we propose that nuclear (or electron) spins in a trapped molecule
would be well isolated from the environment and the state of each spin can
be measured by means of mechanical detection of magnetic resonance (MD of
MR). With modern MR techniques$^{2,3}$, a single trapped molecule amounts
to an appealing piece of hardware for experimental quantum computation. In
the context of MR spectroscopy, a molecular trap promises the ultimate
sensitivity for single spin detection and an unprecedented spectral
resolution as well.

Nuclear spin systems are good for quantum computation because they are
intrinsically isolated from their environment and spin-spin coupling in
a natural spin lattice enables universal quantum logic operations$^{4,5}$.
Quantum computation based on a single molecule or nanocrystal does not
encounter the problem of exponential decay in signal-to-noise ratio
suffered by solution NMR$^6$, provided that the state of each spin can be
directly measured. Unfortunately it seems very difficult to realize the
nanoprobe scheme for detection and dynamical control of single nuclear
spins$^7$. Indeed, scientists have been pursuing the goal of single-spin
detection using various methods, including optical spectroscopy on single
molecules$^{8,9}$ and the very promising magnetic resonance force 
microscopy (MRFM)$^{10-12}$. However detecting single nuclear spins with
MRFM requires high-quality cantilever and extreme magnetic field gradient
which are still out of the reach of today's technology. Besides, with the
sample fixed on a solid, MRFM may not be suitable for NMR quantum
computation due to the decoherence problem. By contrast, a single molecule
or nanoparticle suspended in an electromagnetic or optical trap 
constitutes a harmonic oscillator with small inertia and ultra-sensitive
``spring'' as well as often negligible damping. A tiny force exerted on a
single spin is readily detectable. Better isolation in a molecular trap is
an added bonus to quantum computation.

The extensively used Paul and Penning traps$^{13,14}$ are suitable for
storing charged molecules or particles. Actually small molecular ions
even charged micro-particles have already been stored and studied in
laboratory$^{13,15,16}$. MR study can be conveniently incorporated into
the Penning trap which uses a strong uniform magnetic field for trapping.
Consider a singly charged molecule containing $N\sim 10^2$ nuclei stored
in the Penning trap with a magnetic field $B\sim 5$ T, its mass $M\sim
10Nm_p$ where $m_p=1.67\times 10^{-27}$ kg is the proton rest mass, if
with a nearly spherical shape its radius $R\sim 2.52N^{1/3}a_0$ where
$4a_0=2.12\times 10^{-10}$ m approximates the typical separation between
adjacent nuclei, then the rotational inertia $I\sim 2MR^2/5\sim 25.4N^{5/3}
m_pa_0^2$. Using the trapping parameters for a proton in Ref.14 (Table
II), the translational motion of the molecular ion will be characterized
by $\nu_c\sim 76$ kHz, $\nu_z\sim 318$ kHz, which are the cyclotron and 
axial frequencies respectively. Here one has $\nu_c<\nu_z$, which
indicates the inadequate radial confinement. Another potential, such as a
Paul-type RF (radio frequency) trap, is required to confine the radial
motion. One is then working on a hybrid ion trap. Assuming thermal energy
$\sim 4.2$ K, the ion undergoes classical motion with velocity $v\sim 8.3$
m/sec, occupies an orbit with diameter $\sim 30$ $\mu$m in the $xy$ plane
and oscillates along the $z$-axis with amplitude $A=v/2\pi\nu_z\sim 4$
$\mu$m. The molecular rotation is hardly quantum with the angular speed
$\omega\sim 2.13\times 10^{10}$ sec$^{-1}$. The molecular vibration is of
quantum nature because the typical phonon frequency $\nu_p\sim C/R\sim
10^{12}$ Hz amounts to higher energy than $4.2$ K, here $C\sim 10^3$ m/sec
is an effective sound speed. With vanishing magnetic field inhomogeneity
and negligible relativistic coupling to electric fields, nuclear spins are
entirely isolated from the translational motion and other external
perturbations. Due to the large mismatch between $\nu_p$ and the spin
precession frequency $\nu_s\sim 2\mu_NB/h\sim 76$ MHz, being $\mu_N=5.05
\times 10^{-27}$ J/T the nuclear magneton, lattice vibration mediated
interactions would be weak because they can only take place via
multi-phonon processes and the fast molecular rotation ought average out
the dipole-dipole coupling by the effect of motional narrowing$^{3,17}$.
The numerical values are sufficient to establish the feasibility of
storing heavy molecular ions in electromagnetic traps and isolating
nuclear spins from the environment at the same time. If stronger
confinement is applied along the radial direction by, for instance, a
filament electrode exerting radial electric field, heavy ions with higher
energy may be trapped and probably better at preserving spin coherence.
Besides, there is the possibility of applying external force to drive
molecular rotation in favor of preserving quantum coherence.

To measure a specified spin using MD of MR, one applies an inhomogeneous 
magnetic field and an RF field in resonance to the spin Larmor precession
such that the target spin flips up and down with the frequency $\nu_z$ and
exerts a resonant force on the $z$-axial harmonic oscillator, in the same
spirit as MRFM$^{10-12}$. However the molecular oscillator is far more
sensitive than a mechanical cantilever. The effective ``spring constant''
is extremely small, $k=M(2\pi\nu_z)^2\sim 6.68\times 10^{-12}$ N/m, and
the $Q$ value is so high, $Q>10^9$ (see Ref.14), that damping can be just
neglected. Assuming a field gradient $\partial B/\partial z\sim 100$ T/m,
a typical nuclear moment experiences a force $f\sim\mu_N(\partial
B/\partial z)\sim 5.05\times 10^{-25}$ N which elongates the ``spring'' by
$f/k\sim 7.56\times 10^{-14}$ m. The force is tiny, but mechanical
resonance will significantly amplify its effect. Depending upon $0$ or
$\pi$ phase difference, the tiny force does work $\sim\pm 4fA$ to the
oscillator within each cycle. Then after about $10^7$ cycles, taking time
$t\sim 30$ sec, the resonant force could increase or decrease the
amplitude of the $z$-axial ion oscillation by $\Delta A\sim(4f/k)\nu_zt
\sim 3$ $\mu$m which is more than observable! However, it may impose some
difficulty to flip the nuclear spin at the high rate of $318$ kHz which
requires a strong RF field on the order of hundred gauss. It seems more
appropriate to apply an oscillating magnetic gradient than to flip the
spin so quickly. Then one has to find a way to eliminate the effect of
other spins. For example, one may periodically, but not so quickly, flip
the target spin and simultaneously invert the magnetic field gradient such
that in a complete cycle of spin flip the total contribution due to other
spins vanishes while the contribution from the target spin grows steadily.

Apart from ion traps, a single molecule or nanocrystal embedded in a
micron-sized particle can be trapped in vacuum by means of optical 
levitation using merely a laser beam$^{18-20}$. For a micro particle with
diameter $2R\sim 1$ $\mu$m, the typical phonon frequency $C/R$ is on the
order of GHz still much higher than the nuclear Larmor frequency. The
large rotational inertia may render the particle rotation too slow at
thermal energy, however with virtually no rotational damping at vacuum the
particle may be driven by some intentionally applied torque to rotate at
an angular frequency up to $1$ GHz$^{19,20}$. Again nuclear spins in such
a particle are well isolated from external perturbations and the
dipole-dipole coupling among them can be effectively washed out by the
fast driven rotation. The force sensitivity of an optical trap is extremely
high so that mechanical isolation and optical field stabilization are
necessary or a feedback system should be equipped$^{19}$. Quote from
Ref.19 for an ultra-sensitive example: ``a particle levitated at the focus
of a beam having a $70$ $\mu$m focal spot diameter moves $10$ $\mu$m in
height for a force change or gravity change of about one part in $10^6$ of
the particle's weight''. If a $0.1$ $\mu$m$^3$ sized particle with mass
$M\sim 2\times 10^{-16}$ kg being levitated, the effective ``spring
constant'' turns to be $k\sim 2\times 10^{-16}$ N/m and the frequency of
harmonic oscillation $\nu_z\sim 0.16$ Hz. In a magnetic field gradient
$\partial B/\partial z\sim 500$ T/m generated from a millimeter sized
ferromagnetic particle$^{11}$, the force exerted on a single proton comes
to $f=\mu_p(\partial B/\partial z)\sim 7\times 10^{-24}$ N which displaces
the particle by $f/k\sim 35$ nm. The displacement is readily detectable by
various methods used in measuring the cantilever deflection in atomic
force microscopy$^{21}$, or one may spend some longer time to accumulate
the resonant oscillation and be confident of neglecting the
damping$^{18}$.

The numerical estimation conclusively suggests a new type of NMR
spectroscopy for single molecules stored in currently used
electromagnetic or optical traps, which are subject to further
improvement and optimization. This will extend NMR applications to
probing physical and chemical phenomena at the molecular scale. Some
questions, such as spin dynamics and relaxation in isolated molecules,
may only be addressed with single molecule NMR. In particular, using
molecular traps it may become reality to determine molecular structure
by means of nuclear spin detection, which is of great importance$^{12}$.
Although the spatial resolution may be spoiled by molecular rotation and
translational motion of the target molecule, different nuclei may be
distinguished according to their characteristic gyromagnetic ratios and
chemical environments. Moreover, high spatial resolution can be retained
by inverting positions of the molecule and the source of inhomogeneous
magnetic field, that is, fixing the target molecule but suspending in the
trap a micro ferromagnetic ball which generates a strong inhomogeneous
field$^{22}$. Of course one of the most important and exciting application
of molecular traps goes to spin-resonance quantum computation, provided,
however, that quantum decoherence can be really suppressed to a
sufficiently low rate.

It has been established that two-bit quantum gates such as exclusive-OR
along with one-bit phase shift are sufficient to carry out any unitary
operation$^{23}$. In the context of NMR spectroscopy, the so-called
unitary quantum gates are in effect routinely realized as coherent spin
manipulations$^{24,25}$. If complete knowledge of spin-spin coupling and
the initial spin configuration in a stored molecule are available, its
spin lattice constitutes a quantum processor which is conveniently 
described by the model of quantum cellular automaton, where quantum
state swapping between any two spin sites is a basic and important  
operation$^{4}$. For the trapped molecule to be really qualified as
a QC, it is further assumed that a special nucleus with its uniquely
recognizable gyromagnetic ratio is identified, such that under certain RF
excitation only the special spin is in resonance and flips up and down
periodically while all other spins are out of resonance thus remain
intact. The special nucleus will serve as the read/write port for the
computer, since its spin state can be uniquely measured using MD of MR.
The source of inhomogeneous magnetic field is necessarily variable in
order to coordinate coherent quantum operation and decoherent quantum
measurement. During coherent operation the inhomogeneous magnetic field
must be completely switched off to avoid decoherence, while switching on
the magnetic inhomogeneity for quantum measurement will lead to strong
interaction between nuclear spins and the macroscopic measurement device
which in turn results in the intentional spin state reduction. The
principle of quantum measurement with MD of MR is based on a variant
Stern-Gerlach effect$^{26}$: in the presence of the RF and inhomogeneous
magnetic fields, the wave packet of a spin particle splits according to
spin up or down in the reference frame rotating synchronously with the RF
field, and the harmonic oscillator of the molecular trap experiences
positive or negative driving force correspondingly. Depending upon the
initial phase ($0$ or $\pi$) of the oscillator, the up or down spin state
will be entangled with an enhanced or suppressed harmonic oscillation.
Actually, in the inhomogeneous magnetic field, each spin exerts a static
force except the read/write spin which contributes an oscillating force
on/to the harmonic oscillator, due to the selective RF excitation.
Therefore different spin configurations are strongly entangled with
different equilibrium positions of the harmonic oscillator, the presence
of magnetic inhomogeneity immediately projects the QC onto a classical
state. Once that happens, the classical spin configuration stays stable,
each spin adiabatically follows its local field without longitudinal 
relaxation$^{17}$ despite perturbations due to translational motion and
molecular rotation, except for the case of resonant perturbation which
should be carefully avoided. Given access to the read/write spin, any
other spin can be read and preset by quantum state swapping$^{4,5}$.
Measuring the whole spin lattice effectively initializes the QC, which
thereafter undergoes unitary evolution with the inhomogeneous magnetic
field completely switched off. Upon completion of unitary transformation,
switching on the magnetic inhomogeneity leads to wave reduction of the QC
and the classical state is readily observed using MD of MR. Obviously, the
NMR QC of molecular trap does not share the problem of exponential decay
in signal-to-noise ratio while enjoys the great immunity to decoherence
with solution NMR computation$^{6,25}$. There seems to be neither
principle nor technical limitation in scaling up NMR quantum computation
using molecular traps, if decoherence is indeed not a problem.

Finally it is worth mentioning that molecular traps are suitable to study
single electron spins as well, and magnetic force detection should perform
better there simply because the magnetic moment gets $10^3$ times larger.
Moreover, if the electron spin system in a trapped molecule or nanoparticle
can also be made robust against decoherence, then a QC based on electron
spin resonance will operate much faster than its NMR counterpart.

\end{document}